\documentclass[a4paper,12pt]{article}
\usepackage[ansinew]{inputenc}
\usepackage[T1]{fontenc}
\usepackage[english]{babel}
\usepackage{latexsym}
\usepackage{amsmath}
\usepackage{amssymb}
\usepackage[dvips]{graphicx}
\usepackage{epsfig}
\usepackage{subfigure}
\newcommand{\mysection}[1]    
	{                   
	\section{#1}
	 
	}

\newcommand{\mc}{\mathcal}

\newcommand{\eps}{\varepsilon}
\newcommand{\ol}{\overline}
\newcommand{\nl}{\newline}

\newcommand{\beq}{\begin{equation}}
\newcommand{\eeq}{\end{equation}}
\newcommand{\beqa}{\begin{eqnarray}}
\newcommand{\eeqa}{\end{eqnarray}}
\newcommand{\beqan}{\begin{eqnarray*}}
\newcommand{\eeqan}{\end{eqnarray*}}
\newcommand{\nn}{\nonumber}

\newcommand{\bc}{\begin{center}}
\newcommand{\ec}{\end{center}}

\newcommand{\Rea}{\textrm{Re}}
\newcommand{\Ima}{\textrm{Im}}

\title{Leptogenesis in B-L gauged SUSY with MSSM Higgs sector}
\author{Heidi Kuismanen, Juho Pelto {\small and} Iiro Vilja  \\
 {\small  \textit{Department of Physics and Astronomy, 
University of Turku, 20014 Turku, Finland}}}
\date{\today}

\begin{document}

\maketitle

We study a modified version of a $U(1)_{B-L}$ gauged MSSM that was recently shown to produce a new source of leptogenesis through the CP asymmetry of sneutrinos and antisneutrinos \cite{blsusy}. By taking all superpotential terms and couplings between the MSSM Higgs and $B-L$ scalar sectors into account we find that the model allows a large enough CP violation to explain the observed baryon number to entropy ratio. Monte Carlo analysis shows that a large amount of CP violation can be produced in the decays of the $B-L$ Higgs bosons and that there are two dominating channels that drive CP violation.    
\section{Introduction}
\setcounter{equation}{0}
\setcounter{footnote}{0}

Baryogenesis via leptogenesis \cite{Fukugita:1986hr,luty} is one of the most appealing scenarios of explaining the observed excess of matter over antimatter in the Universe \cite{wmap} indicated by the baryon number to entropy ratio
\beqa
\frac{n_B}{s}=(8.75\pm 0.23)\times 10^{-11}.
\eeqa
In these models right-handed singlet neutrinos, which are also responsible for the nonzero masses of Standard Model (SM) neutrinos \cite{seesaw}, decay to SM leptons and the SM Higgs doublet creating CP asymmetry and violating lepton number. The resulting net lepton number is then converted to baryon number by sphalerons \cite{sphaleron}. The basic picture of leptogenesis has been accommodated into supersymmetric (SUSY) models in various ways \cite{covi,nir,raidal}. In \cite{covi}, the MSSM superpotential is augmented with interactions between singlet right chiral neutrinos and MSSM lepton and Higgs superfields, and a Majorana mass term is also included. With this setup sneutrinos and other superpartners can run in the loop diagrams and the decay products include the superpartners of SM leptons and Higgs bosons as well. In addition, sneutrinos can decay like their fermionic superpartners. References \cite{nir, raidal} consider the effect of soft supersymmetry breaking to leptogenesis (soft leptogenesis). It turns out that a single sneutrino generation can produce the required CP violation and net lepton number as opposed to the standard leptogenesis scenario.

A natural way to extend SM by singlet neutrinos is to gauge $B-L$. Simple extensions of SM with gauged $B-L$ symmetry, i.e. containing the subgroup $U(1)_{B-L}$, must accommodate three right-handed singlet neutrinos to cancel the triangle anomaly $[U(1)_{B-L}]^3$. Gauging $B-L$ within SUSY  models would help understand $R$ parity with the transformation $R=(-1)^{3(B-L)+2S}$ \cite{rparity}. The breaking of $B-L$ can be attributed to additional Higgs fields that carry an even $B-L$ charge and these Higgs bosons also generate the large Majorana masses for the right-handed neutrinos.

A further modified model of supersymmetric leptogenesis was presented in \cite{blsusy} where it was found that MSSM extended with gauged $U(1)_{B-L}$ gives rise to a new source of CP violation. Namely, the new heavy Higgs bosons that spontaneously break the gauged $U(1)_{B-L}$ symmetry can undergo decay into $\tilde N$ and $\tilde N^*$ thus creating an asymmetry between these two. This new asymmetry is converted to conventional lepton asymmetry as the sneutrinos $\tilde N$ decay into MSSM leptons and Higgs bosons and their superpartners. Both resonant leptogenesis \cite{sarkarwf}-\cite{pilafunderwood} and soft leptogenesis arise in this model: CP violation is due to the complex parameters in the soft SUSY breaking sector and the heavy Higgs bosons are degenerate prior to the onset of soft SUSY breaking. After SUSY breaking, the Higgs boson masses receive suppressed contributions leading to a quasidegenerate neutral boson spectrum. 

We elaborate on the model presented in \cite{blsusy} by including in the superpotential and soft SUSY breaking potential the MSSM Higgs sector that couples to the field $S$ and terms such as $S^3$ and $\Delta \ol \Delta$. Thus, the sources of CP violation are the soft terms as well as the terms added to the superpotential, and as a consequence the model does not exhibit soft leptogenesis alone. The mass spectrum for the heavy Higgs sector is assumed to remain quasidegenerate and so resonant leptogenesis pertains to this model. In the present paper, we compute the CP violation parameter and lepton number with these modifications and investigate the allowed parameter regions with Monte Carlo methods. In section 2 we present the model, CP violation is studied in section 3 and the results of the numerical analysis are presented in section 4. The results are discussed in section 5.

\section{The model}
\setcounter{equation}{0}
\setcounter{footnote}{0}

The model is based on the gauge group $SU(3)_C\times SU(2)_L\times U(1)_Y\times U(1)_{B-L}$. The $B-L$ charges for relevant chiral superfields $\ol \Delta, \Delta, S,\ N,\ e^c,\ L,\ Q,\ u^c$ and $d^c $ \\
are $+2,\ -2,\ 0,\ +1,\ +1,\ -1,\ +1/3,\ -1/3$ and $-1/3$, respectively. The $N_i$ fields receive Majorana masses through the vacuum expectation value $\langle \Delta \rangle$ that breaks $B-L$ symmetry. The most general superpotential for these fields reads then as 
\beqa
W^{B-L}&=&\lambda S(\Delta \ol \Delta-M^2)+\frac{1}{2}f_{ij}N_iN_j\Delta\nn\\
&&+Y^{\alpha i}_\nu L_\alpha N_iH_u+\mu H_uH_d+M_1S^2+M_2\Delta\ol\Delta+Y_1S^3+Y_3SH_uH_d,
\eeqa
the D term potential is
\beqa
V_D^{B-L}&=&\frac{1}{8}(g^2+g'^2)(|H_u^0|^2+|H^+_u|^2-|H_d^0|^2-|H^-_d|^2)^2\\
&&+\frac{1}{2}g^2|H^+_uH^{0*}_d+H^0_uH^{-*}_d|^2+2g_B^2(|\Delta|^2-|\ol \Delta|^2)^2\nn
\eeqa
and the soft SUSY breaking potential is
\beqa
V_{soft}^{B-L}&=&\big[b(H_u^+H_d^--H_u^0H_d^0)+a_1S+b_1S^2+b_2\Delta\ol\Delta+c_1S^3+c_2S\Delta\ol\Delta\\
&&+c_3S(H_u^+H_d^--H_u^0H_d^0)+\frac{A_ff_{ij}}{2}\Delta\tilde N_i\tilde N_j\big]+h.c.\nn\\
&&+m_{H_u}^2|H_u|^2+m_{H_d}^2|H_d|^2+m_{S}^2|S|^2+m_{\Delta}^2|\Delta|^2+m_{\ol\Delta}^2|\ol\Delta|^2.\nn
\eeqa
The $SU(2)_L$, $U(1)_Y$ and $U(1)_{B-L}$ gauge couplings are denoted by $g$, $g'$ and $g_B$, respectively. For convenience, we move to unitary gauge by making the transformations
\beqa
\Delta&=&\frac{1}{\sqrt{2}}(|M|+\Delta_0)e^{q_\Delta g_B\Delta'},\nn\\
\ol\Delta&=&\frac{1}{\sqrt{2}}(|M|+\Delta_0)e^{-q_\Delta g_B\Delta'+i\phi_{M}}.
\eeqa
Minimizing the scalar potential $V=V_F^{B-L}+V_D^{B-L}+V_{soft}^{B-L}$ w.r.t. $\Rea(H_u)$, $\Rea(H_d)$, $\Ima(H_u)$ and $\Ima(H_d)$, $\Rea(S)$, $\Ima(S)$, $\Rea(\Delta_0)$, $\Ima(\Delta_0)$ allows us to eliminate the parameters $m_{H_u}^2$, $m_{H_d}^2$, $L_r\equiv\Rea(-\lambda M^2)$, $\Rea(a_1)$, $\Ima(a_1)$, $\Rea(b_2)$ and $L_i\equiv\Ima(-\lambda M^2)$, respectively, and also, $b$ is fixed. We are thus left with the free parameters $M_1$, $M_2$, $Y_1$, $Y_3$, $b_1$, $c_1$, $c_2$, $c_3$, $A_f$, $m_{S}^2$, $m_{\Delta}^2$, $m_{\ol \Delta}^2$ and $\langle S \rangle$. The gauge coupling $g_B$ does not appear in any of the mass eigenvalues or mass eigenvectors so it is not included in the Monte Carlo analysis. 

Our minimization of the scalar potential departs from the corresponding procedure carried out in \cite{blsusy} in that we eliminate some of the soft parameters and randomize $\langle S \rangle $ in the vicinity of 1 TeV. The vacuum expectawtion value (VEV) of $S$ is nonzero, $\langle S \rangle \neq 0$, at SUSY breaking and the vacuum expectation value has to be of the SUSY breaking scale \cite{blsusy}. Because of the alterations made to the superpotential, $\langle S \rangle$ is nonzero before SUSY breaking and altogether $\langle S \rangle \sim 1$ TeV after soft SUSY breaking. The VEV $\langle S \rangle$ appears in the formulas of $L_i$ and $L_r$ as well as the soft parameters that have been eliminated in the minimization procedure. 

From the scalar potential $V=V_F+V_D+V_{soft}$ we determine the mass matrix in the basis $(\Rea H_u, \Ima H_u, \Rea H_d, \Ima H_d, \Rea S, \Ima S, \Rea \Delta_0, \Ima \Delta_0)$. This computation is described in the Appendix. The expressions for the masses and eigenvectors containing the perturbative corrections are very complicated and finding special limits is nontrivial. By limiting different parameters we would obtain different limits for the masses so there are no unequivocal expressions for the masses or the eigenvectors. The four quasidegenerate heavy $B-L$ bosons of this system decay to sneutrinos and antisneutrinos generating CP violation, which we investigate in the next section.

\section{CP violation}
\setcounter{equation}{0}
\setcounter{footnote}{0}

\begin{figure}[ht]
\centering
\subfigure[]{
\includegraphics[scale=0.8]{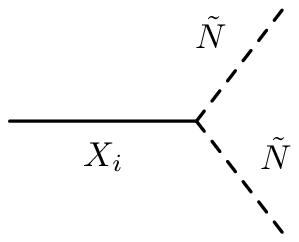}
\label{fig:subfig26}
}
\quad
\subfigure[]{
\includegraphics[scale=0.8]{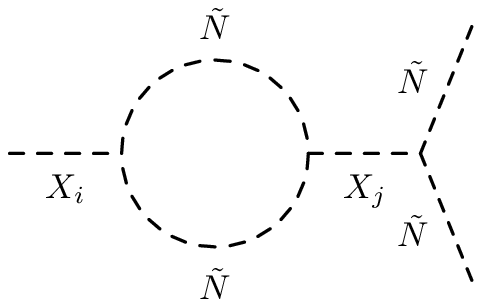}
\label{fig:subfig27}
}
\quad
\subfigure[]{
\includegraphics[scale=0.8]{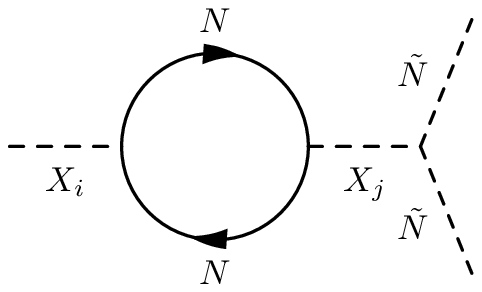}
\label{fig:subfig28}
}
\caption{The relevant tree level and loop diagrams depicting the decay of the heavy mass states $X_i$ to sneutrinos. Sneutrinos or their fermionic partners can run in the loop.}
\label{fig:sneutrinodecays}
\end{figure}

Since we are dealing with a quasidegenerate system in the heavy $B-L$ Higgs bosons, we expect that the main contribution comes from the interference between the tree level decay diagram and mixing diagrams in Fig. \ref{fig:sneutrinodecays}.
To compute the CP violation parameter we also need the corresponding diagrams that produce $\tilde N^*$ pairs. The diagrams of Fig. \ref{fig:sneutrinodecays} arise from the scalar potential part
\beqa \label{scalarint}
V^{(3)}=|\tilde N|^2\sum _iF_{|\tilde N|i}X_i+(\tilde N \tilde N \sum_i F_{\tilde N\tilde N i}X_i+\text{h.c.})
\eeqa
and the Yukawa interaction between the neutrinos and the heavy states $X_i$
\beqa
\mc L_{NNX_i}=NNY_{F_i}X_i+\text{h.c.},
\eeqa
where the couplings $F_{|\tilde N|i}$, $F_{\tilde N\tilde N i}$ and $Y_{F_i}$ are given later. We have restricted ourselves to the case where there is only one neutrino generation $N$. The sneutrino states $\tilde N$ and $\tilde N^*$ mix with each other and form the propagating mass eigenstates
\beqa
\tilde N_+&=&\frac{1}{\sqrt 2}(e^{i\chi}\tilde N+e^{-i\chi}\tilde N^*),\\
\tilde N_-&=&\frac{1}{\sqrt{2}i}(e^{i\chi}\tilde N-e^{-i\chi}\tilde N^*).\nn
\eeqa
Equation (\ref{scalarint}) now becomes
\beqa
V^{(3)}=\sum_i(\tilde N_+^2F_{++i}+\tilde N^2_-F_{--i}+\tilde N_+\tilde N_-F_{+-i})X_i.
\eeqa
The definition for the CP violation parameter
\beqa
\eps=\frac{\sum_i[\Gamma(X_i\rightarrow \tilde N\tilde N)-\ol \Gamma(X_i\rightarrow \tilde N^*\tilde N^*)]}{\sum_i[\Gamma(X_i\rightarrow \tilde N\tilde N)+\ol \Gamma(X_i\rightarrow \tilde N^*\tilde N^*)]}
\eeqa
is employed and after the standard calculation we find
\beqa
\eps&=&4\sum_i\sum_{j\neq i}\frac{M_{X_i}^2-M_{X_j}^2}{(M_{X_i}^2-M_{X_j}^2)^2+\Pi_{jj}^2}\Pi_{ij}\Ima (F_{\tilde N\tilde N i}F_{\tilde N\tilde N j}^*)\frac{1}{16 \pi M_{X_i}}\\
&&\times \Bigg [\sqrt{1-\frac{4M_{\tilde N_+}^2}{M_{X_i}^2}}+\sqrt{1-\frac{4M_{\tilde N_-}^2}{M_{X_i}^2}}+\sqrt{1-2\frac{M_{\tilde N_+}^2+M_{\tilde N_-}^2}{M_{X_i}^2}+\frac{(M_{\tilde N_+}^2-M_{\tilde N_-}^2)^2}{M_{X_i}^4}}\Bigg ]\nn\\
&&\times \Bigg \{ \sum_i\frac{1}{8\pi M_{X_i}}|F_{\tilde N\tilde N i}|^2\Bigg [\sqrt{1-\frac{4M_{\tilde N_+}^2}{M_{X_i}^2}}+\sqrt{1-\frac{4M_{\tilde N_-}^2}{M_{X_i}^2}}\nn\\
&&+\sqrt{1-2\frac{M_{\tilde N_+}^2+M_{\tilde N_-}^2}{M_{X_i}^2}+\frac{(M_{\tilde N_+}^2-M_{\tilde N_-}^2)^2}{M_{X_i}^4}}\Bigg ]\Bigg \}^{-1}\nn
\eeqa 
with the absorptive part of the boson and fermion loops
\beqa \label{absorpt}
\Pi_{ij}&=&\frac{1}{32\pi}(2K_{++}F_{++i}F_{++j}+2K_{--}F_{--i}F_{--j}+K_{+-}F_{+-i}F_{+-j})\\
&&+\frac{1}{16\pi}\sqrt{1-\frac{4M_N^2}{M_{X_i}^2}}\big[(M_{X_i}^2-2M_N^2)(Y^\dagger_FY_F+Y^T_FY^*_F)_{ij}\nn\\
&&-2M_N^2(Y^T_FY_Fe^{-2i\phi_f}+Y^*_FY^\dagger_Fe^{2i\phi_f})_{ij}\big].\nn
\eeqa
The sneutrino squared masses and neutrino mass are
\beqa \label{sneunumasses}
M^2_{\tilde N_{\pm}}&=&\frac{1}{2}|f|^2|M|^2\pm \sqrt{2}|f||M||\lambda\langle S\rangle e^{i\phi}+M_2e^{i\phi}+A^*_f|\\
M_N&=&\frac{1}{\sqrt 2}|f||M|.\nn
\eeqa
The couplings are given by
\beqa \label{couplings}
F_{|\tilde N|i}&=&|f|^2|M|n_{7i},\\
F_{\tilde N\tilde N i}&=&\frac{f}{2\sqrt{2}}\lambda^*\langle S\rangle^* e^{-i\phi}(n_{7i}-in_{8i})+\frac{f}{2\sqrt 2}\lambda^*e^{-i\phi}|M|(n_{5i}-in_{6i})\nn\\
&&+\frac{f}{2\sqrt 2}M_2^*e^{-i\phi}(n_{7i}-in_{8i})+\frac{fA_f}{2\sqrt 2}(n_{7i}+in_{8i}),\nn\\
Y_{F_i}&=&\frac{1}{2 \sqrt 2}f(n_{7i}+in_{8i}),\nn\\
F_{++i}&=&\frac{1}{2}(F_{|\tilde N|i}+e^{-2i\chi}F_{\tilde N\tilde N i}+e^{2i\chi}F_{\tilde N\tilde N i}^*),\nn\\
F_{--i}&=&\frac{1}{2}(F_{|\tilde N|i}-e^{-2i\chi}F_{\tilde N\tilde N i}-e^{2i\chi}F_{\tilde N\tilde N i}^*),\nn\\
F_{+-i}&=&i(e^{-2i\chi}F_{\tilde N\tilde N i}-e^{2i\chi}F^*_{\tilde N\tilde N i}),\nn\\
\chi&=&\frac{1}{2}{\text{arg}}\bigg (\frac{f\lambda^*\langle S\rangle^* e^{-i\phi}+fM_2^*e^{-i\phi}+fA_f}{2\sqrt 2}\bigg).\nn
\eeqa
The coefficients $n_{ij}$ relate the fields $(\Rea H_u, \Ima H_u, \Rea H_d, \Ima H_d, \Rea S, \Ima S, \Rea \Delta_0, \Ima \Delta_0)$ to the mass eigenstates $X_i$ as shown in the Appendix.

Our result for the fermionic part in the second line of the absorptive part of the loop (\ref{absorpt}) is slightly different from that in \cite{blsusy}. We have checked our result with the help of the optical theorem that suggests (\ref{absorpt}) is correct. The correction $-2M_N^2(Y^\dagger_FY_F+Y^T_FY^*_F)$ is of the same order as $-2M_N^2(Y^T_FY_Fe^{-2i\phi_f}+Y^*_FY^\dagger_Fe^{2i\phi_f})$ so the modification is not significant in magnitude compared to the dominating term $M_{X_i}^2(Y^\dagger_FY_F+Y^T_FY^*_F)$.

In the couplings (\ref{couplings}) the combination $\lambda \langle S\rangle +M_2$ appears. This sum equals the VEV of $S$ in \cite{blsusy} if we minimize $V_F$ w.r.t. $S$ before SUSY breaking and keep the dominating terms $\sim |M|^2$. The resulting VEV of $S$ is $\langle S \rangle =M_2/\lambda$ and after soft SUSY breaking $\langle S\rangle$ receives further corrections from the soft potential. This correction $-(c_2^*|M|^2e^{-i\phi}/2+a_1^*)/(|\lambda|^2|M|^2)$ to our $\langle S\rangle $ is the result they found in \cite{blsusy} for $\langle S\rangle$ by keeping the largest terms $\sim |M|^2$ in the $S$ derivative of the scalar potential.

The source for CP violation does not solely lie in the soft SUSY breaking sector in our model because the complex couplings $F_{\tilde N\tilde N i}$ include complex parameters from the superpotential as well. These new contributions come from the heavy $B-L$ sector as well as the coupling part between the $B-L$ and MSSM Higgs sectors, $Y_3SH_uH_d$. Also, the soft sector is modified and, in particular, the scalar $B-L$ and scalar MSSM Higgs sectors are again coupled via the term $c_3SH_uH_d$.

\section{Numerical results}
\setcounter{equation}{0}
\setcounter{footnote}{0}

In determining the viable parameter regions we impose some conditions. First, the excess baryon number created by $\tilde N$ decays is given by \cite{raidal, blsusy}
\beqa
\frac{n_B}{s}\simeq -8.6\times 10^{-4}\eta \eps,
\eeqa
where the washout factor $\eta$ can be $\sim 0.1$ at most \cite{blsusy, Hubble rate, kappafactor, leptogenesislecture}. Thus, $\eps \sim -10^{-6}$ is required. Second, the heavy Higgs decay rates must be smaller than the expansion rate of the Universe. Finally, we want to ensure that the system is quasidegenerate and demand that the heavy Higgs boson mass differences are a few orders of magnitude smaller than the heavy Higgs masses, $M_{X_i}$, themselves. Also, negative mass squared values for the Higgs bosons and sneutrinos may arise and we have to filter these out as well on physical grounds.

\begin{figure}[ht]
\centering
\subfigure[]{
\includegraphics[scale=0.8]{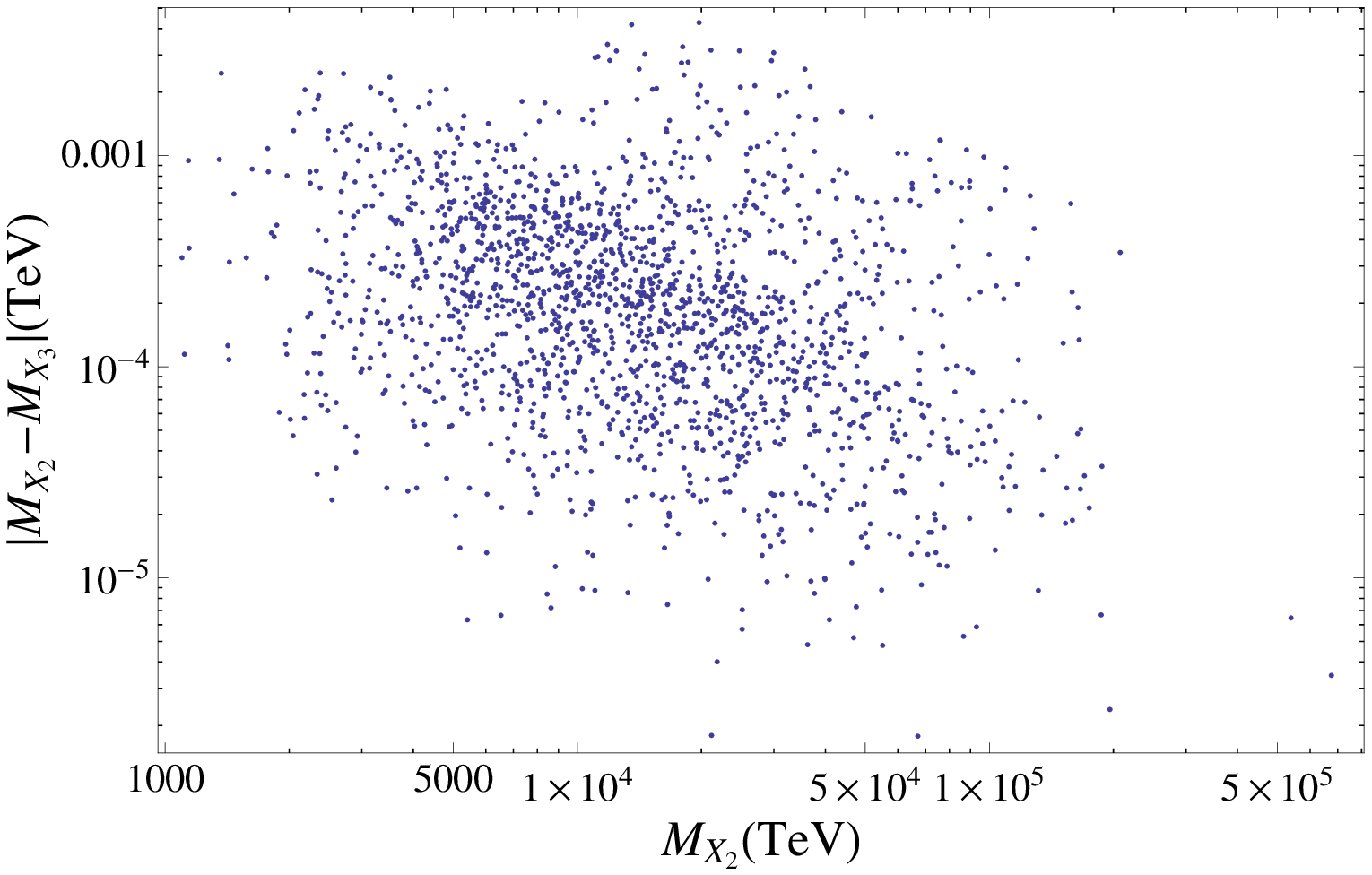}
\label{fig:subfig1}
}
\\
\subfigure[]{
\includegraphics[scale=0.8]{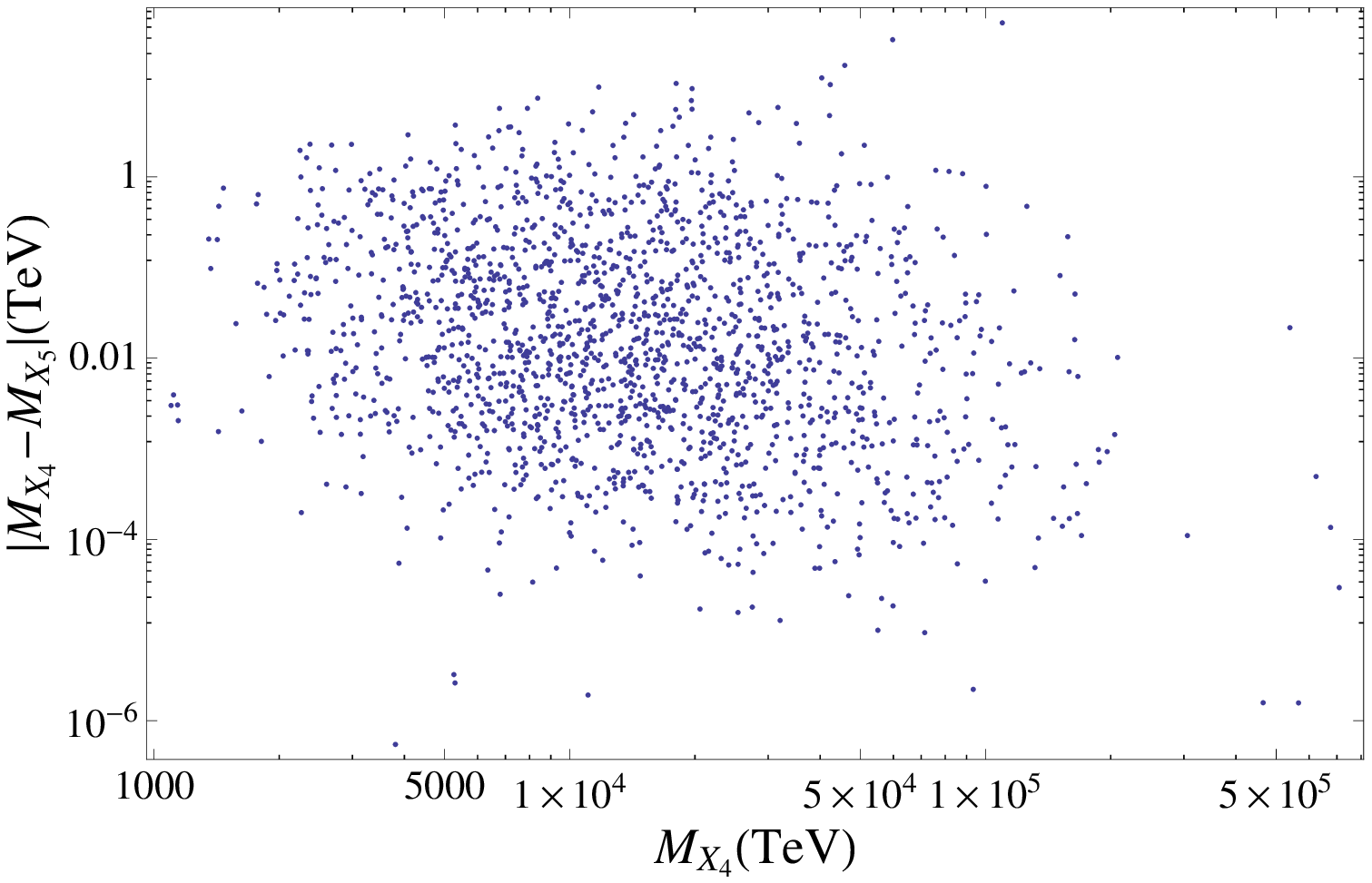}
\label{fig:subfig2}
}
\caption{The mass difference $|M_{X_2}-M_{X_3}|$ versus $M_{X_2}$ in Fig. \ref{fig:subfig1} and $|M_{X_4}-M_{X_5}|$ versus $M_{X_4}$ in Fig. \ref{fig:subfig2}. The difference $|M_{X_2}-M_{X_3}|$ is a few orders of magnitude smaller than $|M_{X_4}-M_{X_5}|$ and other mass differences. $|M_{X_2}-M_{X_3}|$ occurs in the amplitude of $X_2\rightarrow X_3\rightarrow \tilde N \tilde N$ which then dominates the CP violation parameter.}
\label{fig:massdiff}

\end{figure}

\begin{figure}[ht]
\centering
\includegraphics[scale=0.8]{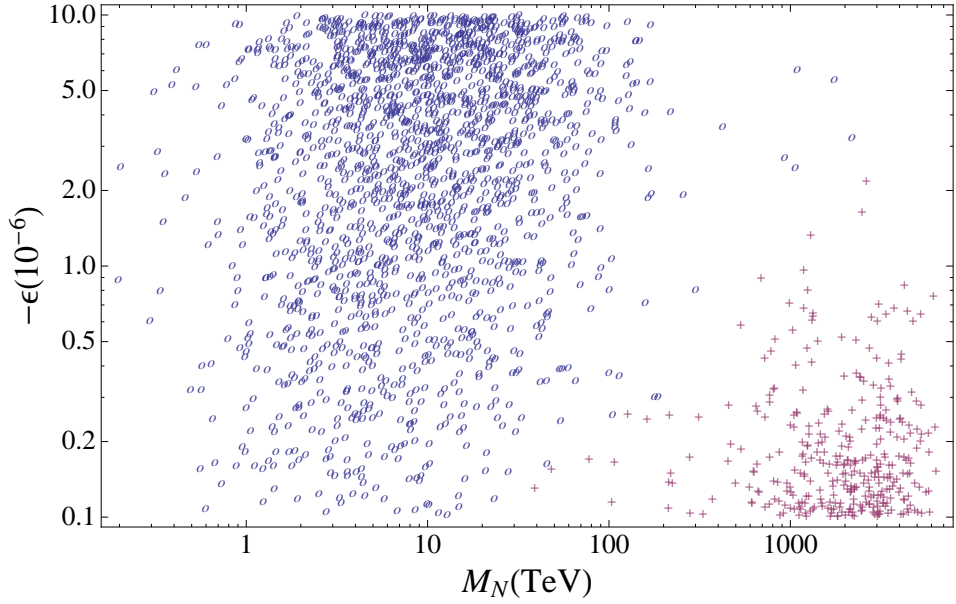}
\caption{CP violation parameter versus neutrino mass $M_N$. Both cases \cite{blsusy} (crosses) and the modified model (circles) are shown. The CP violation parameter $|\eps|$ tends to the higher end compared to \cite{blsusy}. The sneutrino masses $M_{\tilde N_\pm}$ spread around the $M_N$ values by the splitting due to soft SUSY breaking (\ref{sneunumasses}) satisfying the hierarchy $M_{\tilde N_-}<M_N<M_{\tilde N_+}$.}
\label{fig:epssneumass}
\end{figure}

\begin{figure}[ht]
\centering
\subfigure[]{
\includegraphics[scale=0.8]{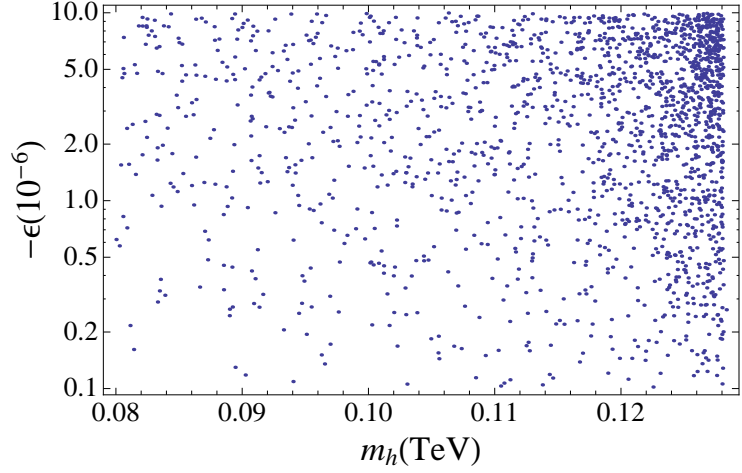}
\label{fig:subfig7}
} 
\quad
\subfigure[]{
\includegraphics[scale=0.8]{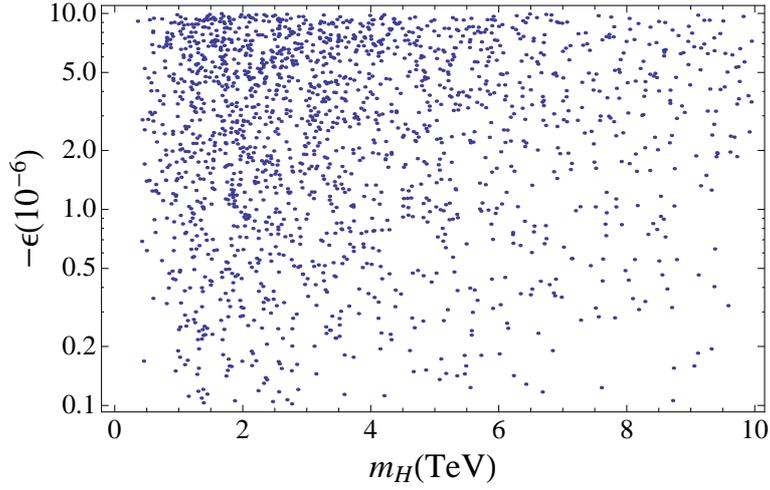}
\label{fig:subfig8}
}
\caption{The CP violation parameter vs $m_h$ in Fig. \ref{fig:subfig7} and $m_H$ in Fig. \ref{fig:subfig8}. The masses $m_h$ and $m_H$ contain corrections from the $B-L$ sector on top of the predictions of MSSM.}
\label{fig:higgsplots}
\end{figure}

\begin{figure}[ht]
\centering
\includegraphics[scale=0.8]{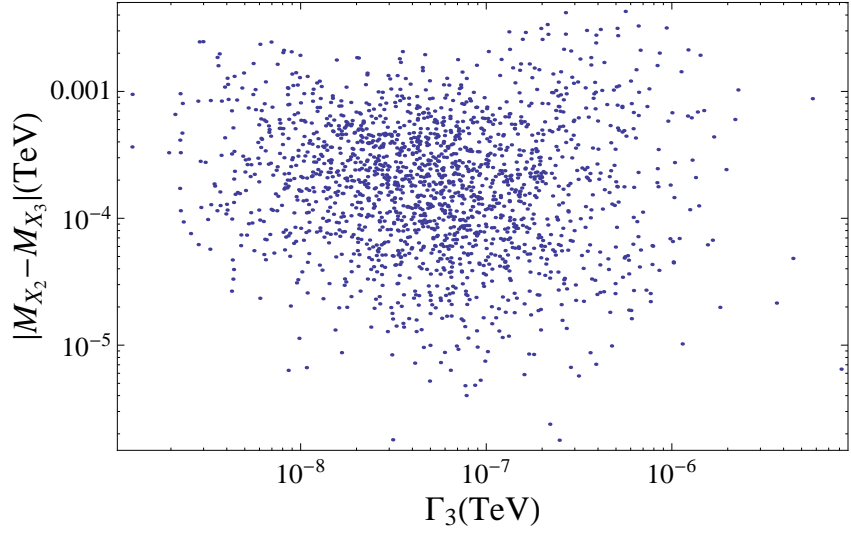}
\caption{The mass difference $|M_{X_2}-M_{X_3}|$ versus the total decay rate of particle $X_3$. This plot illustrates how the situation relates to the resonance condition $|M_{X_2}-M_{X_3}|\sim \Gamma$.}
\label{resonance}
\end{figure}

\begin{figure}[ht]
\centering
\includegraphics[scale=0.7]{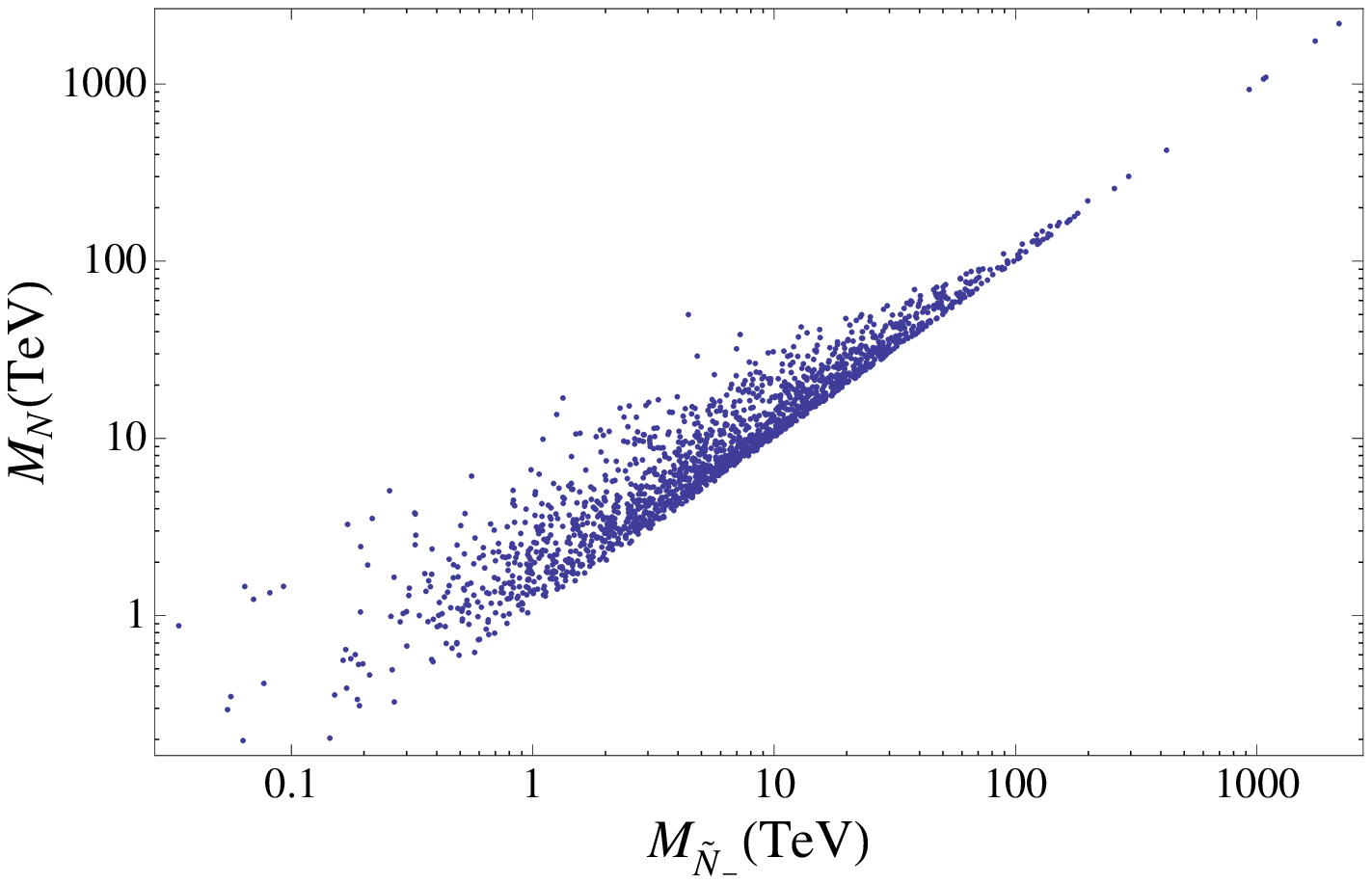}
\caption{Sneutrino mass $M_{\tilde N_-}$ vs neutrino mass $M_N$. The mass differences are of the SUSY breaking scale set by the soft parameters in Table \ref{parameterscan}. }
\label{sneuvsnu}
\end{figure}

By performing Monte Carlo analysis on the system with the new superpotential parameters and soft parameters in Table \ref{parameterscan}  
\begin{table}
\caption{The parameters and their scanned values.}
\label{parameterscan}
\begin{center}
\begin{tabular}{|c|c|} 
\hline
Parameter & Scanned values \\
\hline
$|M_{1,2}|$ & 0.1-10.0 TeV \\
\hline
$|Y_1|$ & 0.01-10 \\
\hline
$|Y_3|$ & $10^{-6}-10^{-3}$ \\
\hline
$|b_1|$ & 0.01-10 TeV$^2$ \\
\hline
$|c_{1,2}|$ & 0.01-10.0 TeV \\
\hline
$|c_3|$ & $10^{-6}-10^{-3}$ TeV \\
\hline
$|A_f|$ & 0.1-10.0 TeV \\
\hline 
$m_S^2,\ m_{\Delta,\ol\Delta}^2$ & 0.1-10.0 TeV$^2$ \\
\hline
$|\lambda|$ & 0.01-0.1 \\
\hline
$|M|$ & $10^4-10^7$ TeV \\
\hline
$|f|$ & $10^{-5}-10^{-3}$ \\
\hline  
$|\langle S\rangle |$ & 0.1-10.0 TeV \\
\hline
$|\mu|$ & 0.1-1.0 TeV \\
\hline
\end{tabular}
\end{center}
\end{table}
\nl and the conditions
\beqa \label{conditions}
-10^{-5}&<&\eps  <-10^{-7},\\  
\Gamma_i &\lesssim& 1.7\sqrt{g_*}\frac{M_{X_i}^2}{M_{Pl}},\nn\\
|M_{X_i}-M_{X_j}|&\lesssim& \frac{M_{X_k}}{1000}\nn
\eeqa
we obtain the plots (each with around 1800 points) in Figs. \ref{fig:massdiff}-\ref{sneuvsnu}. The scan ranges of the soft parameters shown in Table \ref{parameterscan} set the SUSY breaking scale in the vicinity of 1 TeV and it may vary up to 10 TeV. The parameter ranges given in Table \ref{parameterscan} all allow for sufficient CP violation and the corresponding plots thus show a uniform distribution of points. On the other hand, the plots with masses and CP violation show more structure. The third line in (\ref{conditions}) is an {\it ad hoc} assumption made to ensure the quasidegenerate nature of the heavy $B-L$ Higgs boson masses.

It is in fact the first condition on $\eps$ of the set (\ref{conditions}) that most heavily restricts the parameter sets. The heavy Higgs bosons have masses in the range $\sim 10^3-5\times 10^5$ TeV, Fig. \ref{fig:massdiff}. Fig. \ref{fig:massdiff} also shows that in reality there are no points ruled out that would widen the gap between the large masses $M_{X_i}$ beyond the third condition in (\ref{conditions}). Without the CP violation parameter condition the allowed heavy Higgs masses would span uniformly the whole interval from $10^2$ to $10^6$ TeV. Also, the sneutrino and neutrino masses would span the interval of roughly $10^{-1}-10^4$ TeV without any restrictions. When the conditions (\ref{conditions}) are imposed, the allowed range for sneutrino masses is $\sim 0.1-100$ TeV with some points also in the region above $\sim 10^3$ TeV while the neutrino mass lies in the interval $\sim 1-100$ TeV also with some points in the region $\sim 1000$ TeV, Fig. \ref{fig:epssneumass}. The occurrence of points in the larger mass region is due to the behavior of $\eps$ versus sneutrino and neutrino masses. The CP violation parameter $|\eps|$ becomes larger than the allowed interval in (\ref{conditions}) when {\it{e.g.}} $M_{\tilde N_\pm}\gtrsim 10^3$ TeV. Once the sneutrino masses attain values $\sim 10^3$ TeV and higher, $|\eps|$ is reduced and some points pass the first condition of (\ref{conditions}). Similar behavior is observed in the neutrino masses as closer to 1000 TeV some points appear. The Higgs masses $m_h$, $m_A$ and $m_H$ sit near 130 GeV and $1-10$ TeV, respectively, with $m_A<m_H$, Fig. \ref{fig:higgsplots}. Especially the lightest Higgs boson has a mass in the vicinity of 130 GeV after (\ref{conditions}) are imposed. Thus, our setup seems to favor the lower end of the estimated mass range of the lightest Higgs boson. (For a review of the Higgs boson masses in SM and MSSM, see \cite{higgsmass}.)

One remarkable difference to \cite{blsusy} is the fact that $|\eps|$ tends to be too large instead of too small. In fact, $|\eps|$ tends to be close to unity in our model with heavy Higgs masses $\sim 10^5-10^6$ TeV, sneutrino masses $\sim 10^2-10^3$ TeV, neutrino mass $\sim 10^2-10^3$ TeV. Also there seems to be a lower limit to the scale of the CP violation parameter, $|\eps|\gtrsim 10^{-7}$. The channels making the largest contribution to $\eps$ are $X_2\rightarrow \tilde N\tilde N$ via $X_3$ and $X_3\rightarrow \tilde N \tilde N$ via $X_2$. This is due to the 1-2 orders of magnitude smaller mass difference $|M_{X_2}-M_{X_3}|$, Fig. \ref{fig:subfig1}, compared to the other differences which are essentially similar to Fig. \ref{fig:subfig2}. The reason why $|M_{X_2}-M_{X_3}|$ is a few orders of magnitude smaller than the other mass differences is not clear-cut due to the complexity of the mass formulas. This mass difference comes closest to the resonance conditions $|M_{X_2}-M_{X_3}|\sim \Gamma$ \cite{pilafunderwood} although this limit is not reached, see Fig. \ref{resonance}.

To compare our setup with \cite{blsusy} we performed Monte Carlo on the model in \cite{blsusy} with our definition of $\eps$ and found that the phase space is significantly more constrained without the MSSM Higgs sector and other interactions introduced into the superpotential. The yield of points that satisfy (\ref{conditions}) is the order of $10^{-4}$ or less in \cite{blsusy} whereas the extended model has a success rate of the order of a few percent. In \cite{blsusy}, $|\eps|$ takes values many orders of magnitude below $10^{-7}$ and attaining the correct magnitude becomes the threshold. Thus, including the MSSM Higgs sector and other allowed terms radically alters the characteristics of the $B-L$ gauged MSSM. Some differences arise also in sneutrino mass $M_{\tilde N_-}$, and consequently $M_{\tilde N_+}$ and $M_N$, Fig. \ref{fig:epssneumass}. The 1-100 TeV range values are favored in the extended model in contrast to the $\sim$ 1000 TeV and above values in \cite{blsusy}. The sneutrino and neutrino masses are of the same order of magnitude as shown in Fig. \ref{sneuvsnu}, with differences determined by the soft parameters in Table \ref{parameterscan} that set the SUSY breaking scale. 

The minimization procedure on the scalar potential leads to expressions for $L_i$ and $L_r$ and for the soft parameters $\Rea(a_1)$, $\Ima (a_1)$ and $\Rea (b_2)$ where $\Ima (Y_3)$ is located in the denominator and does not cancel. This $1/\Ima (Y_3)$ dependence survives to the mass matrix perturbation $\delta M_{BL}^2$ (\ref{perturbations}) and so the perturbative masses and eigenvectors also have $\sim 1/\Ima (Y_3)$ dependence. By reducing $Y_3$ all heavy Higgs mass differences become smaller but they do not approach the values given by the model in \cite{blsusy}. 


Moreover, we have checked that the use of perturbation theory in diagonalizing the 8$\times$8 Higgs boson squared mass matrix is legitimate because the perturbative corrections to the boson masses are relatively very small, namely $|M_{X_i}-m_{0i}|\lesssim 10$ TeV and for the lightest MSSM Higgs boson $|M_{X_7}-m_{07}|\lesssim 0.01$ TeV. These differences are obviously many orders of magnitude smaller than the masses themselves. 

The scans for $M_1$, $M_2$, $Y_3$ and $c_3$ have been restricted. We have limited $Y_3$ and $c_3$ to very small values because these directly couple the MSSM and $B-L$ Higgs sectors. This would mean that the MSSM Higgs masses, especially $m_h$, could receive large $\sim 1$ TeV corrections. Expanding the range for $Y_3$ and $c_3$ up to $\sim 1$ and $\sim 1$ TeV, respectively, yields less points that satisfy (\ref{conditions}) compared to the values in Table \ref{parameterscan} so it is justified to stick to the lower end of the range. Allowing $M_1$ and $M_2$ to take values up to $\sim M$ makes the use of perturbation theory no longer valid and the VEVs of $\Delta$, $\ol \Delta$ and $S$ would be significantly altered. It is likely that $\langle S \rangle$ would no longer be $\sim 1$ TeV and instead would increase many orders of magnitude. This situation is not considered in our analysis.  

Overall, the phase space of the model is rather complex and no single parameter from the superpotential or soft potential dominates the behavior of physical quantities. Thus, obtaining clear relations between physical quantities, {\it {e.g.}} $|M_{X_2}-M_{X_4}|$ versus $M_{X_4}$, is possible only if the parameters in the superpotential and soft potential are fixed to values that satisfy (\ref{conditions}). Some parameters, {\it {e.g.}} $Y_3$, $c_3$, $Y_1$ and $c_1$, can be allowed to fluctuate around the values that pass (\ref{conditions}) but this causes the masses $M_{X_i}$ themselves to vary by very little, $\mc O(10^{-4})$ TeV. Likewise, limiting physical quantities, {\it {e.g.}} $M_N$, instead of the parameters does not constrain the phase space enough to produce meaningful relations depicting the dependence between other physical quantities like masses, mass differences or the CP violation parameter. On the other hand, the Monte Carlo analysis unveils well the behavior and allowed regions of the MSSM Higgs boson masses corrected by the $B-L$ sector, sneutrino masses, neutrino mass and CP violation parameter. Especially it has revealed that the lightest Higgs boson mass $m_h$ seems to favor the $\sim 130$ GeV region, and that we have two clearly dominant channels and that these channels can produce a large amount of CP violation $\eps \lesssim 1/2$. 


\section{Discussion}
\setcounter{equation}{0}
\setcounter{footnote}{0}

We have found that matter-antimatter asymmetry via an asymmetry between $\tilde N$ and $\tilde N^*$ can be produced in the $B-L$ gauged MSSM with reasonable parameter values. We do not have to resort to finetuning because including the coupling term between the $B-L$ and MSSM Higgs sectors and other allowed terms in the superpotential allows the model to produce a large excess of matter over antimatter. The situation is opposite in \cite{blsusy} where $|\eps|$ tends to be many orders of magnitude below the required lower limit. The success of our model is largely due to the fact that two of the heavy Higgs masses $M_{X_2}$ and $M_{X_3}$ come very close to each other compared to the other masses and subsequently the resonant condition becomes closer for these two particles. The decay channels $X_2\rightarrow X_3^*\rightarrow\tilde N \tilde N$ and $X_3\rightarrow X_2^*\rightarrow \tilde N \tilde N$ thus dominate the CP violation parameter which can become $|\eps|\lesssim 1/2$. 



The results from our model suggest that including the MSSM Higgs sector as well as other allowed interactions may change radically the characteristics of the physical system, in this case the amount of CP violation. Also, many other models that couple the MSSM Higgs bosons to the new Higgs bosons that spontaneously break the symmetry introduced beyond MSSM, see {\it e.g.} \cite{huitu, langacker, langackertoka}, could be worthwhile studying in terms of leptogenesis.

{\bf Acknowledgments} The work of H.K. was supported by Finnish Academy of Science and Letters (Väisälä fund).

\appendix

\setcounter{equation}{0}
\setcounter{footnote}{0}

\section{Diagonalization of the mass squared matrix}

In solving the neutral Higgs boson mass spectrum by diagonalizing an $8\times 8$ matrix we rely on perturbation theory that is familiar from quantum mechanics (see {\it {e.g.}} \cite{ballentine}). The zeroth-order squared mass matrix consists of the MSSM Higgs block situated in the upper left corner and the $B-L$ Higgs sector in the lower right corner. Modifications compared to \cite{blsusy} such as the superpotential coupling $Y_1$ are included in the unperturbed $B-L$ block. The corrections to the squared masses and eigenvectors are computed using the standard formulas 
\beqa \label{perturbations}
\delta V_n&=&\sum _{m \neq n}\frac{(V_m^T\delta M^2_{BL}V_n)V_m}{m_{0n}^2-m_{0m}^2},\\
\delta M_{X_n}^2&=&V_n^T\delta M_{BL}^2V_n,\nn \\
(\delta M^2_{BL})_{ij}&=&\frac{\partial ^2(V_F+V_D+V_{soft})}{\partial \phi_i \partial \phi_j}-M_{BL0}^2,\nn
\eeqa 
where $V_n$ denote the eigenvectors and $m_{0n}^2$ the eigenvalues of the zeroth order squared mass matrix and $M_{BL0}^2$ is given later. The perturbation to the zeroth order squared mass matrix is denoted by $\delta M^2_{BL}$ and it equals the difference between the squared mass matrix, which is derived by differentiation w.r.t. Higgs scalar fields from the scalar potential $V$, and (\ref{zeromass}). The perturbation thus holds various terms including the added couplings to the superpotential $M_1,\ M_2,\ Y_1$ and $Y_3$ and their soft counterparts. We find the mass squared matrix for the scalar bosons in the basis 
\nl $X_i'=(\Rea H_u, \Ima H_u, \Rea H_d, \Ima H_d, \Rea S, \Ima S, \Rea \Delta_0, \Ima \Delta_0)$ and obtain the following decompositions in terms of the mass eigenstates
\beqa
X_i'=\sum_{j=1}^8 n_{ij}X_j
\eeqa
where the coefficients $n_{ij}$ are normalized lengthy expressions including first order perturbative corrections. The masses of the Higgs boson sector are denoted by $M_{X_i}$ and $M_{X_2},...,M_{X_5}$ are the heavy states that arise from the $B-L$ sector. The light Higgs states corresponding to the MSSM Higgses are denoted by $X_6\equiv A$, $X_7\equiv h$ and $X_8\equiv H$. Then the Goldstone boson is massless $M_{X_1}=0$, the lightest Higgs $h$ has mass $M_{X_7}$, $M_{X_6}$ corresponds to the mass of $A$ and the $H$ boson has mass $M_{X_8}$.

The zeroth-order mass squared matrix is in the basis $(\Rea H_u, \Ima H_u, \Rea H_d, \Ima H_d, \\ \Rea S, \Ima S, \Rea \Delta_0, \Ima \Delta_0)$
\beqa \label{zeromass}
M_{BL0}^2=\begin{pmatrix}\frac{b v_d}{v_u}+\frac{(g'^2+g^2)v_u^2}{2} & 0 & -b-\frac{(g'^2+g^2)v_dv_u}{2} & 0 & 0 & 0 & 0 & 0\\0 & \frac{b v_d}{v_u} & 0 & b & 0 & 0 & 0 & 0\\ 
-b-\frac{(g'^2+g^2)v_dv_u}{2} & 0 & \frac{(g'^2+g^2)v_d^2}{2}+\frac{b v_u}{v_d} & 0 & 0 & 0 & 0 & 0\\
0 & b & 0 & \frac{b v_u}{v_d} & 0 & 0 & 0 & 0\\
0 & 0 & 0 & 0 & a_S & b_S & 0 & 0\\
0 & 0 & 0 & 0 & b_S & c_S & 0 & 0\\
0 & 0 & 0 & 0 & 0 & 0 & a_D & b_D\\
0 & 0 & 0 & 0 & 0 & 0 & b_D & c_D \end{pmatrix},
\eeqa
where 
\beqan
a_S&=&6 L_i Y_{1i}+(\Rea(|M|+\langle \Delta_0 \rangle)^2 + \Ima(|M|+\langle \Delta_0 \rangle)^2) (Y_{2i}^2 + Y_{2r}^2)\\
&&+ 3 Y_{1i} (\Rea(|M|+\langle \Delta_0 \rangle)^2 Y_{2i} - \Ima(|M|+\langle \Delta_0 \rangle)^2 Y_{2i} \\
&&+ 2 \Rea(|M|+\langle \Delta_0 \rangle) \Ima(|M|+\langle \Delta_0 \rangle) Y_{2r}) \cos(\phi)\\ 
&&+3 Y_{1i} (-2 \Rea(|M|+\langle \Delta_0 \rangle) \Ima(|M|+\langle \Delta_0 \rangle) Y_{2i} \\
&&+ \Rea(|M|+\langle \Delta_0 \rangle)^2 Y_{2r} - \Ima(|M|+\langle \Delta_0 \rangle)^2 Y_{2r}) \sin(\phi)\\
b_S&=&3 Y_{1r} (2 L_i + (\Rea(|M|+\langle \Delta_0 \rangle)^2 Y_{2i} - \Ima(|M|+\langle \Delta_0 \rangle)^2 Y_{2i} \\
&&+ 2 \Rea(|M|+\langle \Delta_0 \rangle) \Ima(|M|+\langle \Delta_0 \rangle) Y_{2r})\cos(\phi) \\
&&+ (-2 \Rea(|M|+\langle \Delta_0 \rangle) \Ima(|M|+\langle \Delta_0 \rangle) Y_{2i} \\
&&+ \Rea(|M|+\langle \Delta_0 \rangle)^2 Y_{2r} - \Ima(|M|+\langle \Delta_0 \rangle)^2 Y_{2r}) \sin(\phi))\\
c_S&=&-6 L_i Y_{1i} + (\Rea(|M|+\langle \Delta_0 \rangle)^2 + \Ima(|M|+\langle \Delta_0 \rangle)^2) (Y_{2i}^2 + Y_{2r}^2) \\
&&-3 Y_{1i} (\Rea(|M|+\langle \Delta_0 \rangle)^2 Y_{2i} - \Ima(|M|+\langle \Delta_0 \rangle)^2 Y_{2i} \\
&&+ 2 \Rea(|M|+\langle \Delta_0 \rangle) \Ima(|M|+\langle \Delta_0 \rangle) Y_{2r}) \cos(\phi) \\
&&+3 Y_{1i} (2 \Rea(|M|+\langle \Delta_0 \rangle) \Ima(|M|+\langle \Delta_0 \rangle) Y_{2i} \\
&&- \Rea(|M|+\langle \Delta_0 \rangle)^2 Y_{2r} + \Ima(|M|+\langle \Delta_0 \rangle)^2 Y_{2r}) \sin(\phi)\\
a_D&=&\frac{1}{4} ((5 \Rea(|M|+\langle \Delta_0 \rangle)^2 + 3 \Ima(|M|+\langle \Delta_0 \rangle)^2) (Y_{2i}^2 + Y_{2r}^2) \\
&&+4 L_i Y_{2i} \cos(\phi) + (\Ima(|M|+\langle \Delta_0 \rangle) (-Y_{2i} + Y_{2r}) \\
&&+ \Rea(|M|+\langle \Delta_0 \rangle) (Y_{2i} + Y_{2r}))(\Rea(|M|+\langle \Delta_0 \rangle)(Y_{2i}-Y_{2r}) \\
&&+ \Ima(|M|+\langle \Delta_0 \rangle) (Y_{2i} + Y_{2r})) \cos(2 \phi) + 4 L_i Y_{2r} \sin(\phi) \\
&&+2 (-\Ima(|M|+\langle \Delta_0 \rangle) Y_{2i} + \Rea(|M|+\langle \Delta_0 \rangle) Y_{2r}) (\Rea(|M|+\langle \Delta_0 \rangle) Y_{2i} \\
&&+ \Ima(|M|+\langle \Delta_0 \rangle) Y_{2r}) \sin(2 \phi))\\
b_D&=&\frac{1}{2} (Y_{2r}\cos(\phi)-Y_{2i}\sin(\phi))(2 L_i+(\Rea(|M|+\langle \Delta_0 \rangle)^2 Y_{2i}-\Ima(|M|+\langle \Delta_0 \rangle)^2 Y_{2i}\\
&&+2 \Rea(|M|+\langle \Delta_0 \rangle) \Ima(|M|+\langle \Delta_0 \rangle) Y_{2r}) \cos(\phi) \\
&&+ (-2 \Rea(|M|+\langle \Delta_0 \rangle) \Ima(|M|+\langle \Delta_0 \rangle) Y_{2i} \\
&&+ \Rea(|M|+\langle \Delta_0 \rangle)^2 Y_{2r} - \Ima(|M|+\langle \Delta_0 \rangle)^2 Y_{2r}) \sin(\phi))\\
c_D&=&\frac{1}{4} ((3 \Rea(|M|+\langle \Delta_0 \rangle)^2 + 5 \Ima(|M|+\langle \Delta_0 \rangle)^2) (Y_{2i}^2 + Y_{2r}^2) \\
&&-4 L_i Y_{2i} \cos(\phi) - (\Ima(|M|+\langle \Delta_0 \rangle) (-Y_{2i} + Y_{2r}) \\
&&+ \Rea(|M|+\langle \Delta_0 \rangle) (Y_{2i} + Y_{2r}))(\Rea(|M|+\langle \Delta_0 \rangle)(Y_{2i}-Y_{2r}) \\
&&+\Ima(|M|+\langle \Delta_0 \rangle) (Y_{2i} + Y_{2r})) \cos(2 \phi) - 4 L_i Y_{2r} \sin(\phi)\\ 
&&+2(\Ima(|M|+\langle \Delta_0 \rangle) Y_{2i} - \Rea(|M|+\langle \Delta_0 \rangle) Y_{2r}) (\Rea(|M|+\langle \Delta_0 \rangle) Y_{2i} \\
&&+ \Ima(|M|+\langle \Delta_0 \rangle) Y_{2r}) \sin(2 \phi)).
\eeqan
In the above formualas, $Y_2\equiv \lambda$. The subscripts $i$ and $r$ denote the imaginary and real parts, respectively. The eigenvectors of (\ref{zeromass}) are 
\beqa
V_1&=&\frac{1}{\sqrt{v_u^2 + v_d^2}}(0, -v_u, 0, v_d, 0, 0, 0, 0)^{\text T}\nn\\
V_2&=& \frac{1}{\sqrt{(a_D - \sqrt{4 b_D^2 + (a_D - c_D)^2} - c_D)^2 + 4 b_D^2}} \nn\\
&&\times (0, 0, 0, 0, 0, 0, a_D - \sqrt{4 b_D^2 + (a_D - c_D)^2} - c_D, 2 b_D)^{\text T}\nn\\
V_3&=&\frac{1}{\sqrt{(a_D + \sqrt{4 b_D^2 + (a_D - c_D)^2} - c_D)^2 + 4 b_D^2}} \nn\\
&&\times (0, 0, 0, 0,0, 0, a_D + \sqrt{4 b_D^2 + (a_D - c_D)^2} - c_D, 2 b_D)^{\text T}\nn\\
V_4&=&\frac{1}{\sqrt{(a_S - \sqrt{4 b_S^2 + (a_S - c_S)^2} - c_S)^2 + 4 b_S^2}} \nn\\
&&\times (0, 0, 0, 0,a_S - \sqrt{4 b_S^2 + (a_S - c_S)^2} - c_S, 2 b_S, 0, 0)^{\text T}\\
V_5&=&\frac{1}{\sqrt{(a_S + \sqrt{4 b_S^2 + (a_S - c_S)^2} - c_S)^2 + 4 b_S^2}} \nn\\
&&\times (0, 0, 0, 0,a_S + \sqrt{4 b_S^2 + (a_S - c_S)^2} - c_S, 2 b_S, 0, 0)^{\text T}\nn\\
V_6&=&\frac{1}{\sqrt{v_u^2 + v_d^2}} (0, v_d, 0, v_u, 0, 0, 0, 0)^{\text T}\nn\\
V_7&=&(\cos(\alpha), 0, -\sin(\alpha), 0, 0, 0, 0, 0)^{\text T}\nn\\
V_8&=&(\sin(\alpha), 0, \cos(\alpha), 0, 0, 0, 0, 0)^{\text T}\nn
\eeqa
with
\beqa
\alpha&=&\arctan((((g'^2+g^2)v_d(v_d-v_u)v_u(v_d+v_u)+2b(-v_d^2+v_u^2)\nn\\
&&-(-8b(g'^2+g^2)v_dv_u(v_d^2-v_u^2)^2+(2b\nn\\
&&+(g'^2+g^2)v_dv_u)^2(v_d^2+v_u^2)^2)^{1/2})/(((g'^2+g^2)v_d(v_d-v_u)v_u(v_d+v_u)\nn\\
&&+2b(-v_d^2+v_u^2)-(-8b(g'^2+g^2)v_dv_u(v_d^2-v_u^2)^2+(2b\nn\\
&&+(g'^2+g^2)v_dv_u)^2(v_d^2+v_u^2)^2)^{1/2})^2\nn\\
&&+(2v_dv_u(2b+(g'^2+g^2)v_dv_u))^2)^{1/2})/((2v_dv_u(2b\nn\\
&&+(g'^2+g^2)v_dv_u))/(((g'^2+g^2)v_d(v_d-v_u)v_u(v_d+v_u)\nn\\
&&+2b(-v_d^2+v_u^2)-(-8b(g'^2+g^2)v_dv_u(v_d^2-v_u^2)^2+(2b\nn\\
&&+(g'^2+g^2)v_dv_u)^2(v_d^2+v_u^2)^2)^{1/2})^2\nn\\
&&+(2v_dv_u(2b+(g'^2+g^2)v_dv_u))^2)^{1/2}))
\eeqa
and the squared masses are
\beqa
m_{01}^2&=&0\nn\\
m_{02}^2&=&\frac{1}{2} (a_D - \sqrt{4 b_D^2 + (a_D - c_D)^2} + c_D)\nn\\
m_{03}^2&=&\frac{1}{2} (a_D + \sqrt{4 b_D^2 + (a_D - c_D)^2} + c_D)\nn\\
m_{04}^2&=&\frac{1}{2} (a_S - \sqrt{4 b_S^2 + (a_S - c_S)^2} + c_S)\\
m_{05}^2&=&\frac{1}{2} (a_S + \sqrt{4 b_S^2 + (a_S - c_S)^2} + c_S)\nn\\
m_{06}^2&=&b (\frac{v_d}{v_u} + \frac{v_u}{v_d})\nn\\
m_{07}^2&=&\frac{1}{4 v_d v_u} (2 b (v_d^2 + v_u^2) + (g'^2 + g^2) v_d v_u (v_d^2 + v_u^2) \nn\\
&&-\sqrt{-8 b (g'^2 + g^2) v_d v_u (v_d^2 - v_u^2)^2 + (2 b + (g'^2 + g^2) v_d v_u)^2 (v_d^2 + v_u^2)^2})\nn\\
m_{08}^2&=&\frac{1}{4 v_d v_u} (2 b (v_d^2 + v_u^2) + (g'^2 + g^2) v_d v_u (v_d^2 + v_u^2) \nn\\
&&+ \sqrt{-8 b (g'^2 + g^2) v_d v_u (v_d^2 - v_u^2)^2 + (2 b + (g'^2 + g^2) v_d v_u)^2 (v_d^2 + v_u^2)^2}).\nn
\eeqa


\begin{thebibliography}{99}

\bibitem{blsusy} K.S. Babu, Yanzhi Meng, Zurab Tavartkiladze, {\textit{New Ways to Leptogenesis with Gauged B-L Symmetry}}, Phys. Lett. B{\bf 681}:37-43 (2009).
\bibitem{Fukugita:1986hr} M. Fukugita and T. Yanagida, {\textit{Baryogenesis Without Grand Unification}} Phys.\ Lett.\  B{\bf 174}:45 (1986).

\bibitem{luty}M. A. Luty, {\textit{Baryogenesis via Leptogenesis}}, {\textit{Phys. Rev.}}  D{\bf 45}:455-465 (1992).

\bibitem{wmap}C. L. Bennett {\it et al.} (WMAP Collaboration), Astrophys. J. Suppl. Ser. {\bf 148}, 1 (2003); D. N. Spergel {\it et al.}, Astrophys. J. Suppl. Ser. {\bf 148}, 175 (2003).
\bibitem{seesaw}P. Minkowski, {\textit{mu --> e gamma at a Rate of One Out of 1-Billion Muon Decays?}}, Phys. Lett. B{\bf 67}:421 (1977); M. Gell-Mann, P. Ramond and R. Slansky in: Supergravity, P. van Nieuwenhuizen, D. Z. Freedman (Eds.), North-Holland, Amsterdam, p. 315 (1979); T. Yanagida, in: Proceedings of the Workshop on the Baryon Number of the Universe and Unified Theories, Tsukuba, Japan 13-14 (1979); S. L. Glasgow, NATO Adv. Study Inst. Ser. B Phys. {\bf 59}:687 (1979); R. N. Mohapatra and G. Senjanovic, {\textit{Neutrino Mass and Spontaneous Parity Violation}}, Phys. Rev. Lett. {\bf 44}:912 (1980).
\bibitem{sphaleron}V. A. Kuzmin, V. A. Rubakov and M. E. Shaposhnikov, {\textit{On anomalous electroweak baryon-number non-conservation in the early Universe}}, Phys. Lett. B{\bf 155}, 36 (1985).
\bibitem{covi}L. Covi, E. Roulet, F. Vissani, {\textit{CP violating decays in leptogenesis scenarios}}, Phys. Lett.  B{\bf 384}:169-174 (1996).
\bibitem{nir}Y. Grossman, T. Kashti, Y. Nir, E. Roulet, {\textit{Leptogenesis from Supersymmetry Breaking}}, Phys. Rev. Lett. 91:25 (2003).
\bibitem{raidal}G. D'Ambrosio, G. F. Giudice, M. Raidal, {\textit{Soft leptogenesis}}, Phys. Lett. B{\bf 575}:75 (2003).
\bibitem{rparity}R. N. Mohapatra, {\textit{New Contributions to Neutrinoless Double beta Decay in Supersymmetric Theories}}, Phys. Rev. D{\bf 34}:3457-3461 (1986); A. Font, L. E. Ibanez and F. Quevedo, {\textit{Does Proton Stability Imply the Existence of an Extra Z0?}}, Phys. Lett. B{\bf 228}:79 (1989); S. P. Martin, {\textit{Some simple criteria for gauged R-parity}}, Phys. Rev. D{\bf 46}:2769 (1992); AC. S. Aulakh, A. Melfo and G. Senjanovic, {\textit{Minimal supersymmetric left-right model}}, Phys. Rev. D{\bf 57}:4174 (1998); M. J. Hayashi and A. Murayama, {\textit{Radiative breaking of $SU(2)_R \times U(1)_{B-L}$ gauge symmetry induced by broken N=1 supergravity in a left-right symmetric model}}, Phys. Lett. B{\bf 153}:251 (1985).
\bibitem{sarkarwf}M. Flanz, E. A. Paschos, U. sarkar, J. Weiss, {\textit{Baryogenesis through mixing of heavy Majorana neutrinos}}, Phys. Lett. B{\bf 389}:693-699 (1996).
\bibitem{coviwf}L. Covi, E. Roulet, {\textit{Baryogenesis from mixed particle decays}}, Phys. Lett.  B{\bf 399}:113-118 (1997).
\bibitem{pilaftsiseka}A. Pilaftsis, {\textit{CP violation and baryogenesis due to heavy Majorana neutrinos}}, Phys. Rev.  D{\bf 56}:5431-5451 (1997). 
\bibitem{pilaftsistoka}A. Pilaftsis, {\textit{Heavy Majorana neutrinos and baryogenesis}}, Int. J. Mod. Phys. A14:1811-1858 (1999).
\bibitem{pilafunderwood}A. Pilaftsis, T. E. J. Underwood, {\textit{Resonant leptogenesis}}, Nucl. Phys. B{\bf 692}:303 (2004).

\bibitem{Hubble rate} E.W. Kolb, M. S. Turner, {\textit{The Early Universe}}, Addison-Wesley Publishing Company (1993).
\bibitem{kappafactor}W. Buchmüller, T. Yanagida, {\textit{Quark lepton mass hierarchies and the baryon asymmetry}}, Phys. Lett. {\bf B}445:399-402 (1999).
\bibitem{leptogenesislecture}M.-C. Chen, {\textit{TASI 2006 Lectures on Leptogenesis}}, hep-ph/0703087 (2007).
\bibitem{higgsmass}M. Carena, H. E. Haber, {\textit{Higgs Boson Theory and Phenomenology}}, Prog. Part. Nucl. Phys. {\bf 50}:63 (2003).
\bibitem{huitu}K. Huitu, J. Maalampi, {\it The Higgs sector of a supersymmetric left-right model}, Phys. Lett. B{\bf 344}:217 (1995).
\bibitem{langacker}T. Han, P. Langacker, B. McElrath, {\it Higgs sector in a} $U(1)'$ {\it extension of the minimal supersymmetric standard model}, Phys. Rev. D{\bf 70}:115006 (2004).
\bibitem{langackertoka}J. Kang, P. Langacker, T. Li, {\it Neutrino masses in supersymmetric} $SU(3)_C\times SU(2)_L\times U(1)_Y\times U(1)'$ {\it models}, Phys. Rev. D{\bf 71}:015012 (2005).
\bibitem{ballentine}L. E. Ballentine, {\textit{Quantum Mechanics A Modern Development}}, World Scientific Singapore (1998).

\end{thebibliography}
\end{document}